\documentclass[showpacs,amsmath,amssymb,amsfonts,pra,floatfix,superscriptaddress,twocolumn]{revtex4} 
\usepackage{graphicx}
\usepackage{bm}
\usepackage[german,english]{babel}
\begin{document}

\title{Controlled excitation and resonant acceleration of ultracold few-boson systems by driven interactions in a harmonic trap}

\author{Ioannis Brouzos}
\email{ibrouzos@physnet.uni-hamburg.de}
\affiliation{Zentrum f\"ur Optische Quantentechnologien, Universit\"at Hamburg, Luruper Chaussee 149, 22761 Hamburg, Germany}

\author{Peter Schmelcher}
\email{Peter.Schmelcher@physnet.uni-hamburg.de}
\affiliation{Zentrum f\"ur Optische Quantentechnologien, Universit\"at Hamburg, Luruper Chaussee 149, 22761 Hamburg, Germany}

\date{\today}

\begin{abstract}
We investigate the excitation properties of finite utracold bosonic systems in a one-dimensional harmonic trap with a time-dependent interaction strength. The driving of the interatomic coupling induces excitations of the relative motion exclusively with specific and controllable contributions of momentarily excited many-body states. Mechanisms for selective excitation to few-body analogues of collective modes and acceleration occur in the vicinity of resonances. We study via the few-body spectrum and a Floquet analysis the excitation mechanisms, and the corresponding impact of the driving frequency and strength as well as the initial correlation of the bosonic state. The fundamental case of two atoms is analyzed in detail and forms a key ingredient for the bottom-up understanding of cases with higher atom numbers, thereby examining finite-size corrections to macroscopic collective modes of oscillation. 
\end{abstract}
 \maketitle

\section{Introduction}

Detailed control of both single particle potential landscapes and interparticle interactions is an appealing feature of ultracold atom physics \cite{pethick_smith}. Apart from cooling the atoms to ultra low temperatures, it is in the meanwhile routinely possible to design almost arbitrarily shaped optical and magnetic traps and to tune them with an unprecedented control \cite{bloch}. The interactions among ultracold atoms can be adjusted by exploiting magnetically and optically induced Feshbach resonances \cite{chin}. The dimensionality of the system can be tuned by strongly confining the transverse degrees of freedom leading to quasi one-dimensional traps and dimensionality-specific phenomena like the Tonks Gas \cite{kinoshita,paredes} of impenetrable bosons \cite{girardeau} and its attractive excited state counterpart, the Super Tonks gas \cite{haller}.  For these low dimensional systems an additional tool  to control the interactions are confinement induced resonances \cite{olshanii}. 

In the present work we take advantage of this experimental progress concerning the design of external and most importantly interatomic forces, and consider the effects of a time-dependent oscillating interaction strength in a one-dimensional harmonic trap. Time-dependent driving is usually applied to external traps, providing inspiring effects such as the dynamical control of tunneling \cite{lignier,zenesini}, dynamical localization \cite{eckardt}, photon-assisted tunneling \cite{sias} via a periodic driving of the lattices or the excitation of collective oscillations in a harmonic trap \cite{moritz} to mention only a few. However, investigations considering a time-dependent scattering length, usually referred to as 'Feshbach resonance management' \cite{kevrekidis03} for the mean-field situation, have inspired a lot of research on control of solitons \cite{kevrekidis} or modulational instabilities \cite{adhidary}. Experimental investigations in this direction have been performed recently \cite{donley,bagnato}. The main advantage of the driving of the scattering length compared to other driving modes applied on the external potential for examining collective excitations \cite{stringari,moritz}, is that other species or non-condensate fractions are not affected by the driving of the interaction of one-species \cite{bagnato}. Apart from the harmonic trap, a two mode system with time-varying interaction has been studied with Floquet theory, leading to many-body coherent destruction of tunneling and localization \cite{gong}.

On the other hand, beyond the mean-field regime, there are relatively few works dealing with the time-dependent modulation of the scattering length, addressing mainly the experimental results on the formation of molecules \cite{donley} from a few-body perspective \cite{molmer,blume}. While these works concentrate mainly on the attractive part of the two-body spectrum and on the corresponding bound state, our work focuses exclusively on repulsive interactions. As illustrated in Ref. \onlinecite{molmer} in a harmonic trap the coupling of an excited two-body state to the molecular ground state is very efficient since nearby states are out of resonance, and possess a relatively small coupling to the initial state. For the repulsive case as we will demonstrate, the relevant few-body states reflect the equidistant spectrum of the harmonic trap and therefore lead to a more complex dynamics involving several instantaneous configurations. A recent publication  \cite{petrov} explored the integrability of the system, via a similar model with time-modulated interaction thereby calculating the dynamical structure factor.

In this work, we aim to examine from a few body perspective the effects resulting from a periodic modulation of the repulsive interaction strength, focusing on few-body collective excitations, control of the dynamics and state population, as well as mechanisms of acceleration via resonances. The case of two atoms in an one-dimensional harmonic trap for which the energy spectrum is known analytically for a constant arbitrary strength of the interaction \cite{busch} serves as a starting point of the investigation. The successful experimental preparation of few-body systems with a controllable number of atoms has been  realized in Ref. \cite{friedhelm} and the energies have been measured with high precision \cite{selim}. Since the modulation of the interaction strength affects exclusively the relative motion in this system (the center of mass is decoupled and therefore unaffected), we study explicitly  the internal motion. Firstly the focus is on the frequencies and the driving amplitudes that give rise to a controlable excitation to particular states after preparation in a certain initial state with a specific strength of the interaction. Additionally we examine particular few-body analogues of collective macroscopic modes of breathing oscillations, as well as a resonant acceleration mechanism via multiple excitations. Our analysis of resonances and acceleration modes is supported by calculations of the Floquet spectrum for the effective single degree of freedom of the relative motion within a harmonic trap and an oscillating delta barrier. Going to higher atom numbers we demonstrate similarities and analyze the differences with the basic case of two particles, and compare the results with those of macroscopic calculations, showing finite size effects on the collective modes. All calculations are performed by the numerically exact Multi-Configurational Time-Dependent Hartree method (MCTDH see Appendix), which is especially designed to treat the dynamics of many degrees of freedom under time-dependent modulations.

This article is organized as follows: In Section II we introduce our model, and  in Section III our focus is on the case of two particles  thereby examining the mechanisms of controlable collective excitations to specific states and the influence of parameter changes on them. We investigate in Section IV the acceleration mechanism via multiple excitations and calculate the Floquet spectrum of this case illustrating the underlying mechanism for the appearance of the resonances. An extension of this study to higher atom numbers is performed in Section V concentrating on the analogue of the breathing mode for collective oscillations and finite size corrections. In the last Section VI we summarize our results and provide an outlook.

\section{Modeling the time dependent interaction}

Quasi one-dimensional waveguides can be created by choosing a strongly focused laser field yielding strongly confined transversal directions compared to the longitudinal one. In this way the trap becomes highly anisotropic with the characteristic length for the transversal trapping   $a_{\perp} \equiv \sqrt {\frac{\hbar}{M \omega_{\perp}} }$ much smaller than the longitudinal one ($\omega_{\perp}$ is the frequency of the harmonic confinement in the transversal direction). Consequently the transverse degrees of freedom are energetically frozen as only the ground state is occupied and the  effective 1D interaction strength reads for the case of contact interactions \cite{olshanii}:   
\begin{equation}
\label{eq1gd}
g_{1D}= \frac{2\hbar^2 a_0}{M a^2_{\perp}} (1-\frac{|\zeta(1/2)| a_0}{\sqrt{2} a_{\perp}})^{-1}
\end{equation}
where the free-space s-wave scattering length $a_0$ does not depend on the detailed appearance of the potential for the interatomic interaction, which is then modeled by an effective contact potential. There are two parameters in Eq.\ref{eq1gd} that can be tuned to attain a time-dependent interaction strength: (i) the scattering length $a_0$ via a change of the strength of e.g. a magnetic field $B$ approaching to or departing from a Feshbach resonance --Feshbach resonance management--  as $a_0 = a_{bg} (1 - \frac{\Delta B}{B-B_0})$ where $\Delta B$ and $B_0$ are the width and the position of the resonance, respectively, and $a_{bg}$ is the background scattering, and (ii) the transversal length $a_{\perp}$ by modifying the relevant laser parameters, taking into account the quasi-one dimensional restrictions posed above (see also Ref. \onlinecite{bagnato}).

The one-dimensional $N$-body Hamiltonian with a time dependent coupling $g(t)$ reads:
\begin{equation}
H=\sum_{i=1}^N \frac{p_i^2}{2}+\sum_{i=1}^N \frac{x_i^2}{2}+g(t)\sum_{i<j}\delta(x_i-x_j)
\end{equation}
where we have performed a scaling transformation setting the length scale equal to the longitudinal characteristic oscillator length $a_{\parallel}\equiv \sqrt {\frac{\hbar}{M \omega_{\parallel}} }$ and the energy scale to $\hbar\omega_{\parallel}$, while the scaled interaction strength is  $g=g_{1D} \sqrt {M/\hbar^3\omega_{\parallel}}$. The interaction potential between each pair of particles $i,j$ is represented by the Dirac $\delta$-function. We note that for numerical purposes, a Gaussian with a very small width of the order of the grid spacing is employed. 

Initially, the particles are prepared in the ground state of the harmonic trap with an interaction strength $g_0$. We will then explore the excitation dynamics for a periodic driving of the repulsive interaction strength of the form:
\begin{equation}
\label{gt}
g(t)=g_0 + \Delta g \sin^2(\omega t) 
\end{equation}
where $\Delta g$ is the amplitude of the driving and $\omega$ the driving frequency. The impact of each of these three parameters of the driving law ($g_0,\Delta g,\omega$) will be examined. The reason for the specific choice  $\sin^2(\omega t)$ for the driving is our focus on repulsive interactions, i.e., $g$ should stay positive even for $g_0=0$. Since $\sin^2(\omega)=\frac{1-\cos(2\omega)}{2}$, the above driving law comprises a periodic oscillation with frequency $2\omega$. Investigating purely attractive interactions as done in Ref. \onlinecite{molmer} or alternating between attractive and repulsive interactions such as in Refs. \onlinecite{blume,donley,kevrekidis03} represent interesting but different situations.

\section{Relative Motion of the two atom problem and instantaneous eigenspectrum}
In general, $N$ particles in a harmonic trap with contact interaction represents a separable problem $H=H_R+H_r$. The center of mass  $R=\frac{1}{N} \displaystyle\sum_{i=1}^N x_i$ and  and its conjugate momentum $P=-i \displaystyle\sum_{i=1}^N \frac{\partial}{\partial x_i}$ constitute the center of mass Hamiltonian $H_R \equiv \frac{p_R^2}{2N}+\frac{N R^2}{2}$. The Hamiltonian of the relative motion $H_r$ is in general not subject to further simplifications and cannot be solved analytically. Nevertheless, for the special case of two particles the relative motion ($r=x_1-x_2$) reduces to an effective one-body problem:
\begin{equation}
\label{hamrel}
H_r=p_r^2+\frac{r^2}{4}+g(t)\delta(r)
\end{equation}
The contact interaction affects only the relative motion, leaving unaffected the center of mass. Therefore we focus on the relative motion, which actually represents a one particle problem with a harmonic trap and a delta barrier with oscillating height placed in the center. With the transformation $r=\sqrt{2}x$, $g(t)=\sqrt{2}g'(t)$ we obtain the standard form of the harmonic oscillator Hamiltonian $H'_{r}=\frac{p_x^2}{2}+\frac{x^2}{2}+g'(t)\delta(x)$. $H_r$ defines an analytically solvable eigenvalue equation \cite{busch} in the case of a time-independent parameter $g$, which we will discuss next as it is very important for the understanding of the excitation dynamics of the relative motion. The solutions cover in general the complete interval $g \in[0,\infty)$ and we will refer to them as the instantaneous eigenstates $\phi_n$, where $n=0,1,2...$ are the energy levels for a certain time instant $t_{ref}$. In spite of the existence of these stationary solutions the driven time-dependent problem possesses no closed analytical solution, although a study via the evolution of the coefficients in an expansion with respect to the corresponding instantaneous eigenstates is natural \cite{molmer,blume}.

\begin{figure}
\includegraphics[width=4.2 cm,height=4.2 cm]{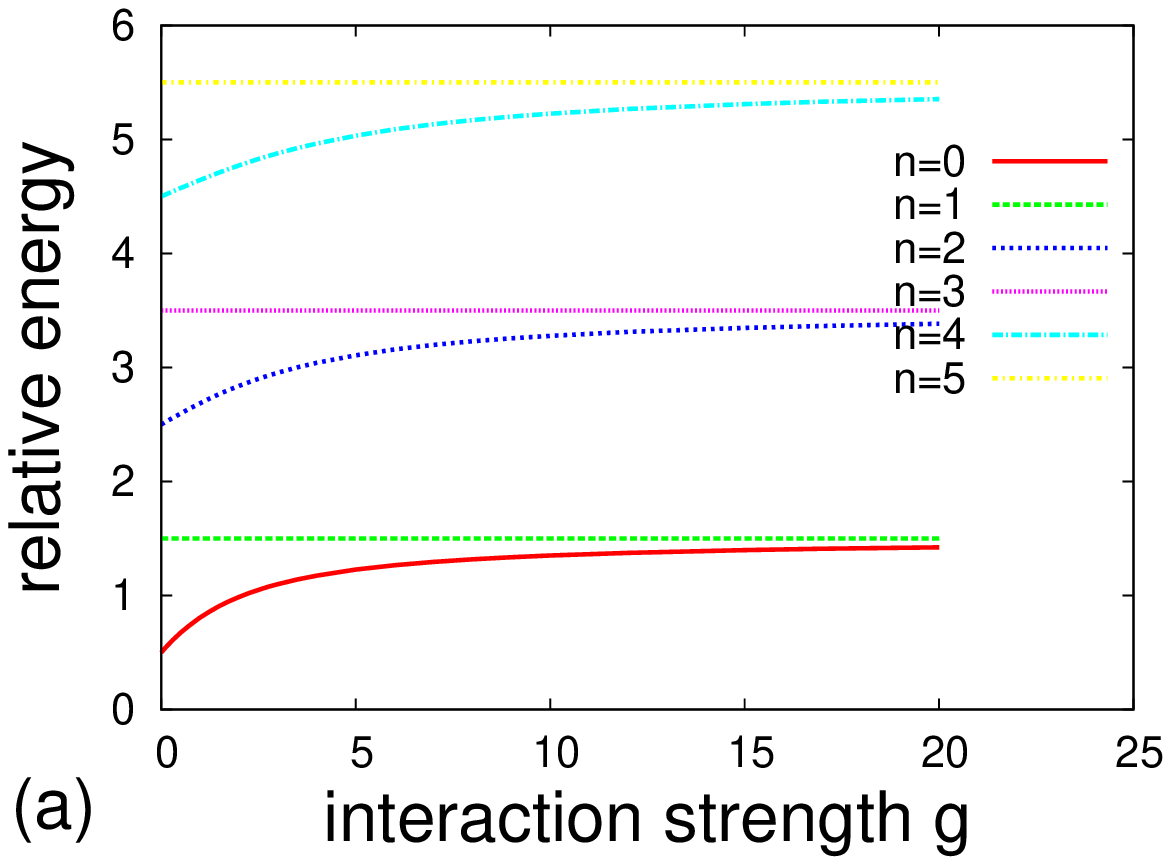}
\includegraphics[width=4.2 cm,height=4.2 cm]{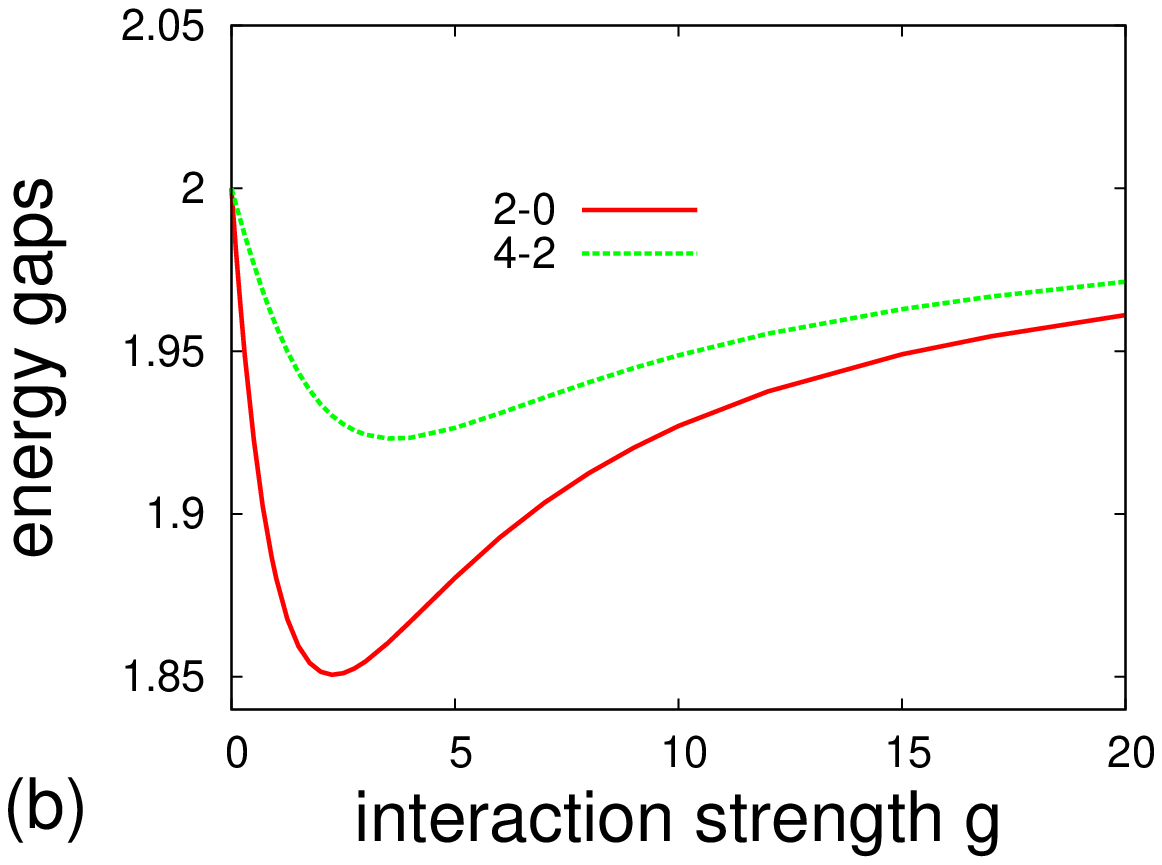}
\caption{(a) The three lowest eigenenergies of the symmetric (0,2,4) and antisymmetric (1,3,5) eigenstates of the relative motion for two particles with increasing interaction strength $g$. (b) The energy difference between the eigenstates of the relative motion $\epsilon_2-\epsilon_0$ and  $\epsilon_4-\epsilon_2$ with increasing $g$.}
\label{fig1}
\end{figure}

In  Fig.~\ref{fig1}(a) we show the lowest lying eigenenergies for $H_r$. For $g=0$ the eigenspectrum of $H_r$ is the harmonic oscillator single-particle spectrum with the equidistant eigenenergies $\epsilon_n=n+1/2$. As $g$ increases the odd levels are unaffected by $\delta(x)$ since they  possess a node at the coordinate origin, while the even states acquire an increasing energy and a dip at $x=0$. Therefore, as we observe in Fig.~\ref{fig1}(a) each even level approaches energetically the next upper odd level forming a doublet spectrum, characteristic for double-well potentials. In the limit where the 'barrier height' is infinite $g \to \infty$  the even levels become degenerate with the odd ones. This limit $g \to \infty$ is the so-called Tonks-Girardeau limit where the bosons are mapped to non-interacting fermions \cite{girardeau}. The Tonks-Girardeau Gas, is one of the most fundamental systems appearing exclusively in one-dimensional many-body systems. The transition to 'fermionization' for two atoms has been recently reported also experimentally \cite{selim}, enhancing the interest on few-body studies, like the present one. 

The energy difference between two even levels with increasing interaction strength shown in Fig.~\ref{fig1}(b) plays also a crucial role for the dynamics. In fact, since the initial preparation is in the ground state which corresponds to an even state of the relative motion, the dynamics can only lead to a population of other even states, and the corresponding energy distance is crucial for the time-evolution. For the two limits of zero interaction and the Tonks-Girardeau gas ($g \to \infty$) the gap between two even parity levels is $|\epsilon_n-\epsilon_{n-2}|_{g=0 \:, \: g\to \infty}=2$. Starting from $g=0$ and increasing $g$ this value slightly decreases [see Fig.~\ref{fig1}(b)], since the states with larger quantum number $n$ possess a lower probability density at $x=0$ and are therefore less affected by the contact interaction.  The slope at $g=0$ reads:
\begin{equation}
\label{atg0}
 \frac{d\epsilon_n}{dg}\Big\vert_{g=0}=\langle \phi_n|\delta(x)|\phi_n \rangle=|\phi_n(0)|^2=\frac{\pi}{n!\Gamma^2\left(\frac{1 - n}{2}\right)}
\end{equation}
The response to a minor increase of  $g$ therefore depends on the value of the harmonic oscillator eigenstates at $x=0$ which decreases with $n$ [see  Fig.~\ref{fig1}(a)]. Therefore at the onset of the interactions the successive even states tend to approach each other energetically. Fig.~\ref{fig1}(b) shows the effect of an increasing coupling strength $g$ on the distance between the first two even levels $\epsilon_0$ and $\epsilon_2$ (red line). The most rapid change of this energy gap is near $g=0$ with a minimum value $\epsilon_2-\epsilon_0 \approx 1.85$ at $g^* \approx 2.2$  and it approaches asymptotically the value $2$ for $g \to \infty$. The slope $\frac{d\epsilon(g)}{dg}$ is decreasing as $g$ increases [see Fig.~\ref{fig1}(a)] and from some value of $g$ on, the energy gap starts to increase again and asymptotically approaches the value of the non-interacting system. Additionally, since  $\frac{d\epsilon_n}{dg}$ decreases with increasing $n$ the deviation from the value $2$ for $\epsilon_2-\epsilon_0$ shown in Fig.~\ref{fig1}(b) (red line) is the largest possible such deviation between two successive (even) states. This is exemplarily shown in the same figure by the distance between the next pair of successive states $e_2$ and $e_4$ (green line). It is obvious that the energy gap $\epsilon_4-\epsilon_2$ is always larger than the $\epsilon_2-\epsilon_0$ gap, and this holds analogously also for gaps between higher lying neighboring states. 
 
The above discussed features of the energy spectrum and the corresponding gaps, possess a crucial impact on the dynamics which we will discuss later.  As $g(t)$ changes with time according to Eq. (\ref{gt}) different regions of $g$-values in Fig. \ref{fig1}  are probed according to the choice of the parameters $g_0$ and $\Delta g$. We might already foresee e.g. that a driving around small $g$-values possesses a greater impact than for a driving for larger $g$-values, since the corresponding slope is larger, subsequently leading to larger energy variations. The equidistance of the spectrum close to the two extreme limits, as well the decrease of the gaps at small to intermediate values of $g$ will also be of great importance concerning a possible resonant behaviour (see sections IV and V). 

We note that Fig.~\ref{fig1} is based on the exact eigenenergies obtained from the equation \cite{busch}:
\begin{equation}
-2^{\frac{3}{2}}\frac{\Gamma(\frac{3-2\epsilon(g)}{4})}{\Gamma(\frac{1-2\epsilon(g)}{4})}=g
\end{equation} 
The corresponding even eigenstates are given analytically in the form of parabolic cylinder functions. For the numerical calculations of the time-dependent evolution of the system we use a regularized delta-function of the form: $\delta_{\sigma}(x)=\frac{\exp(-\frac{x^2}{2\sigma^2})}{\sqrt{2\pi}\sigma}$ with $\sigma=0.05$ which is small enough to catch the real 'delta-like' behavior but also convenient for a numerical grid sampling. Minor numerical deviations on the eigenstates and the quantum evolution stemming from this approximation are unavoidable.

\section{resonant controllable excitation}

The wave packet of the relative motion at $t=0$ corresponds to the ground state of $H_{r}$ in Eq.~(\ref{hamrel}) with $g(0)=g_0$. This an even state and since parity is conserved during the time evolution only even parity states can be occupied, which relates to the bosonic permutation symmetry. In this work we are interested mainly in two quantities:
\begin{itemize}
 \item the population of momentarily eigenstates $\phi_n$ at a certain time $p_n(t_{ref})=|\langle \phi_{n}(t_{ref}) | \psi (t_{ref}) \rangle|^2$, where $\psi(t)$ is the wave function of the relative motion. We use $t_{ref}=\frac{2\pi k}{\omega}$ ($k \in \mathbb{N}_0$) without loss of generality. 
 \item the time evolution of the energy $\epsilon (t)$. We refer only to the relative energy since the center of mass is completely untouched by the change of $g$.  
\end{itemize}

\begin{figure}
\includegraphics[width=4.2 cm,height=4.2 cm]{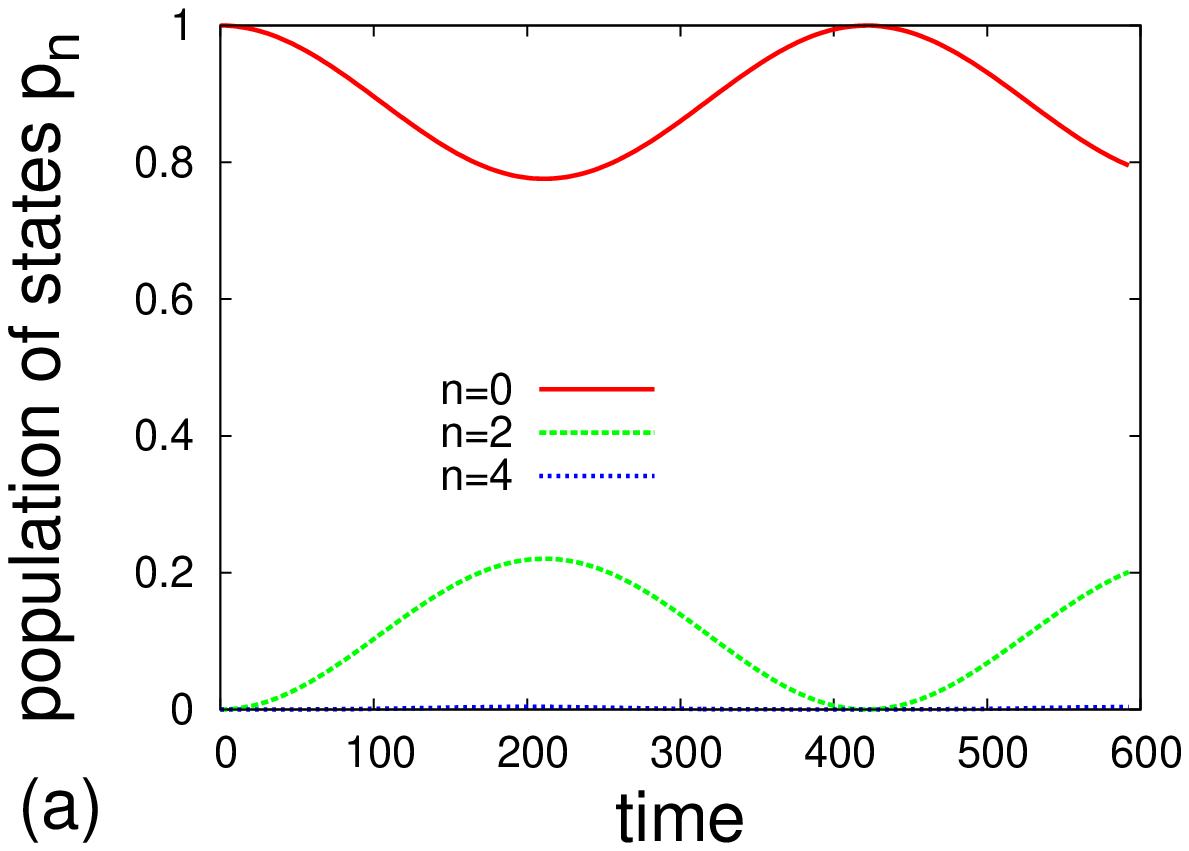}
\includegraphics[width=4.2 cm,height=4.2 cm]{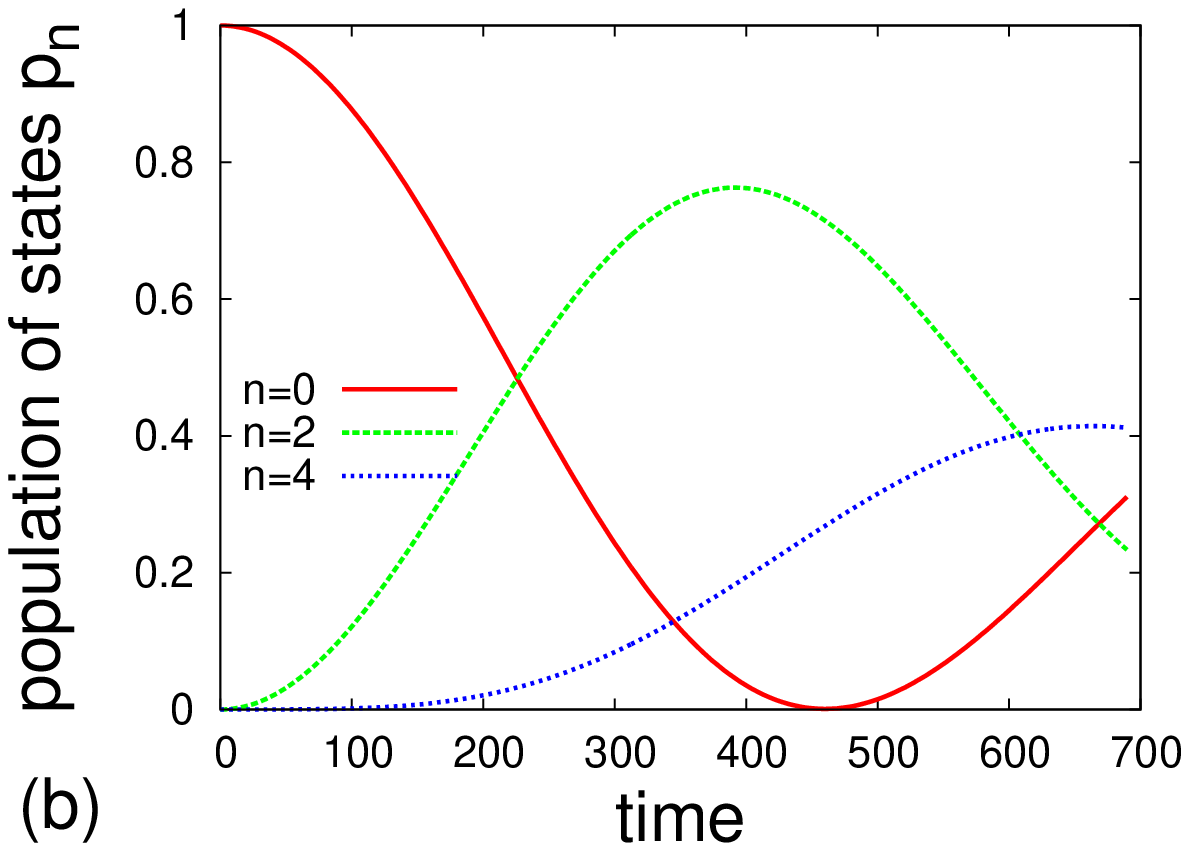}
\includegraphics[width=4.2 cm,height=4.2 cm]{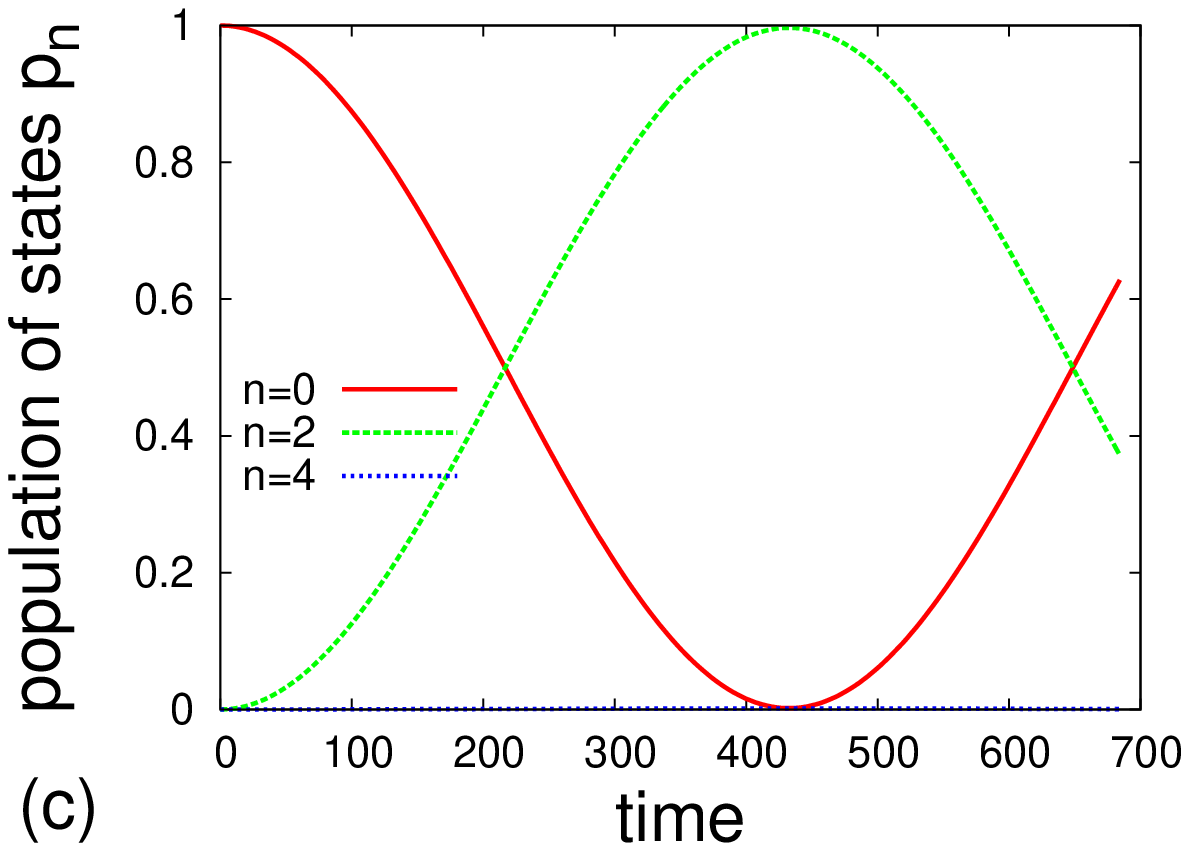}
\includegraphics[width=4.2 cm,height=4.2 cm]{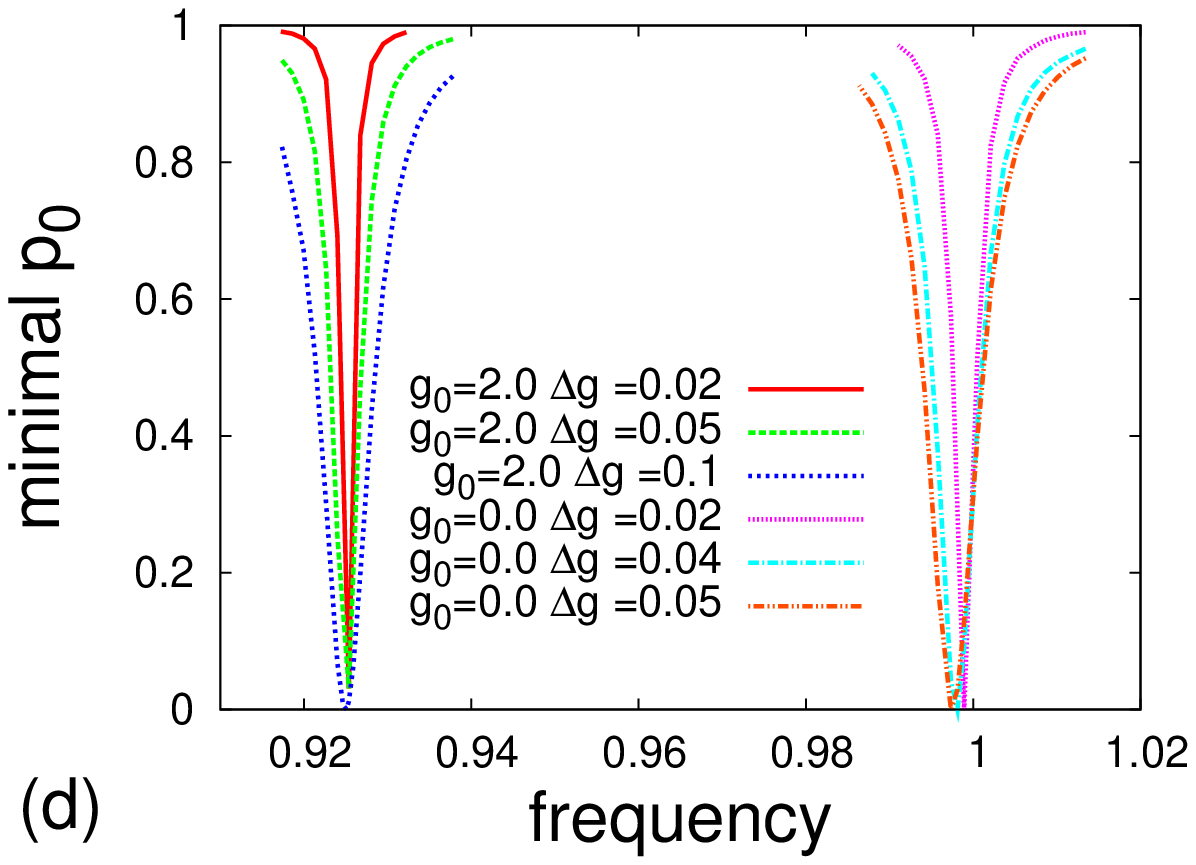}
\caption{Population of instantaneous eigenstates with time variation, for (a) $g_0=0.0$, $\Delta g=0.05$, $\omega=0.99$, (b) $g_0=0.0$, $\Delta g=0.05$, $\omega=0.997$, (c) $g_0=2.0$, $\Delta g=0.1$, $\omega=0.925$. Profile of the resonances via the minimal occupation of the ground state with varying driving frequency. Several cases are shown for different values of  $g_0$ and $\Delta g$.}
\label{fig2}
\end{figure}

We start our investigation in the regime of small driving amplitudes where the excitation dynamics is to a larger extent controllable. The main role in this regime is played by the driving frequency. If this frequency is much lower than the gap between two even states ($\omega<<1$) then we are in the adiabatic regime and the evolution involves only the momentarily ground state $\phi_0$ (with $p_0>0.99$). Approaching the first resonance $\omega=1$ from below induces an excitation to the state $\phi_2$ of the relative motion. This is shown for a typical case in Fig.~\ref{fig2} (a) where $g_0=0$, $\Delta g=0.05$ and $\omega=0.99$. We see that the ground state looses population while the second excited state gains. The next even level $\phi_4$ remains almost unpopulated. For the particular case though where $g_0=0$ (and correspondingly for $g_0 \to \infty$) since the eigenspectrum is initially completely equidistant, an excitation to the next level $\phi_4$ is not prohibited corresponding to a two step process. Therefore we see in Fig ~\ref{fig2} (b) that a bit closer to the resonance ($\omega=0.997$)  the $n=4$ level gains population after the $n=2$ level does so. This is also in general the case for larger amplitudes $\Delta g>0.5$ where multiple excitations are enhanced as we will see later on. Therefore, while the system departs -- in this case completely -- from the ground state, the first collective excitation to state $n=2$ is necessarily combined with a transfer of population to the next level and therefore a controllable excitation exclusively to the second level (complete depopulation of ground state and complete population of the second excited level) is not possible here. It becomes though possible if the initial interaction is stronger. For instance for $g_0=2.0$, $\Delta g=0.1$ we see in Fig.~\ref{fig2} (c) a complete transfer of population to the first excited level. Therefore, the non-harmonicity in the spectrum due to the initial correlations is helpful from the point of view of a controllable state excitation and preparation. Let us note that for a controlable creation of such states, one should choose a small amplitude $\Delta g <0.5$, since larger amplitudes lead easily to multiple occupation of excited states due to the close to equidistant spectrum. Additionally the driving frequency should be carefully tuned to be close to the corresponding energy spacing of the spectrum with $g(0)=g_0$. For this case $\omega=0.925$ --somewhat lower than the resonant frequency for the non-interacting $g_0=0$ situation--  we have a decreased gap in the energy spectrum [see Fig.~\ref{fig1} (b)].

As an overview of the resonances in this regime we present in Fig.~\ref{fig2} (d) the minimal occupancy of the instantaneous ground state $p_0$ as a function of the driving frequency for several small amplitudes of the driving. We observe that far from resonance the minimal $p_0 \approx 1$, so there is hardly any excitation as expected. The frequency where the ground state becomes at a certain time completely unoccupied corresponds to the resonance. The resonant frequency is slightly shifted to lower values for a larger amplitude of the driving which is attributed to the  decrease of the energy gap in the spectrum [see Fig.~\ref{fig1} (b)], as $\Delta g$ covers larger regions of $g$ (for $g_0=0$). We also observe that for $g_0=2.0$ the resonant frequency is shifted to much lower values than for $g_0=0$, approximately corresponding to the energy gap of the levels at $g=2.0$ ($\omega=\frac{\epsilon_2-\epsilon_0}{2} \approx 0.925$). For this case though, small changes in the driving amplitude do not shift significantly the position of the resonance, since the energy spacing in the vicinity of this value of $g_0$ does not change substantially [see Fig.~\ref{fig1} (b)]. However, the most important difference of the two cases for different initial coupling strength $g_0$ is the fact that the case $g_0=0$ leads easily to multiple excitation of higher states since the energy gaps are all close to the same value corresponding to the same resonant frequency, while for larger $g_0$ the energy spacing is not equidistant and therefore a complete controlable transfer to a certain state is possible. 

A comment on the robustness of the initial state preparation is in order here. Apart from being easily excitable to different instantaneous states, the initially non-interacting ensemble is in general more sensitive to the driving of the interaction. Even far from resonance the evolution of this initial state ($g_0=0$), leads to a change in energy of the order of $10 \%$ while for stronger and  particular intermediate interactions as well as close to the fermionization regime $g_0=20.0$ it is ten orders of magnitude lower. This is understandable if one inspects the slopes of the energy curves in Fig ~\ref{fig1} (a), which are much larger for small values of $g$. This could be a signature for the detection of a highly correlated ensembles like the Tonks Girardeau gas, i.e., by studying their response to changes of the interaction strength.  

\begin{figure}
\includegraphics[width=4.2 cm,height=4.2cm]{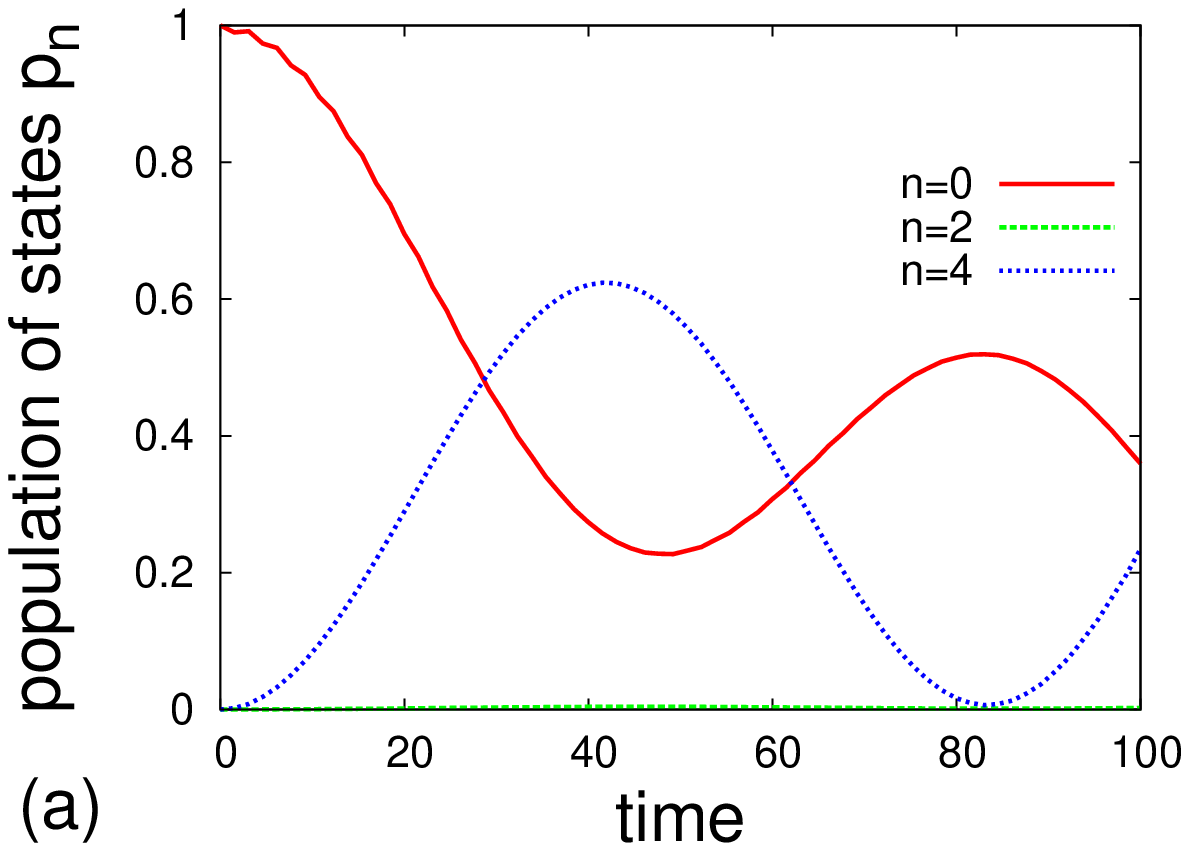}
\includegraphics[width=4.2 cm,height=4.2cm]{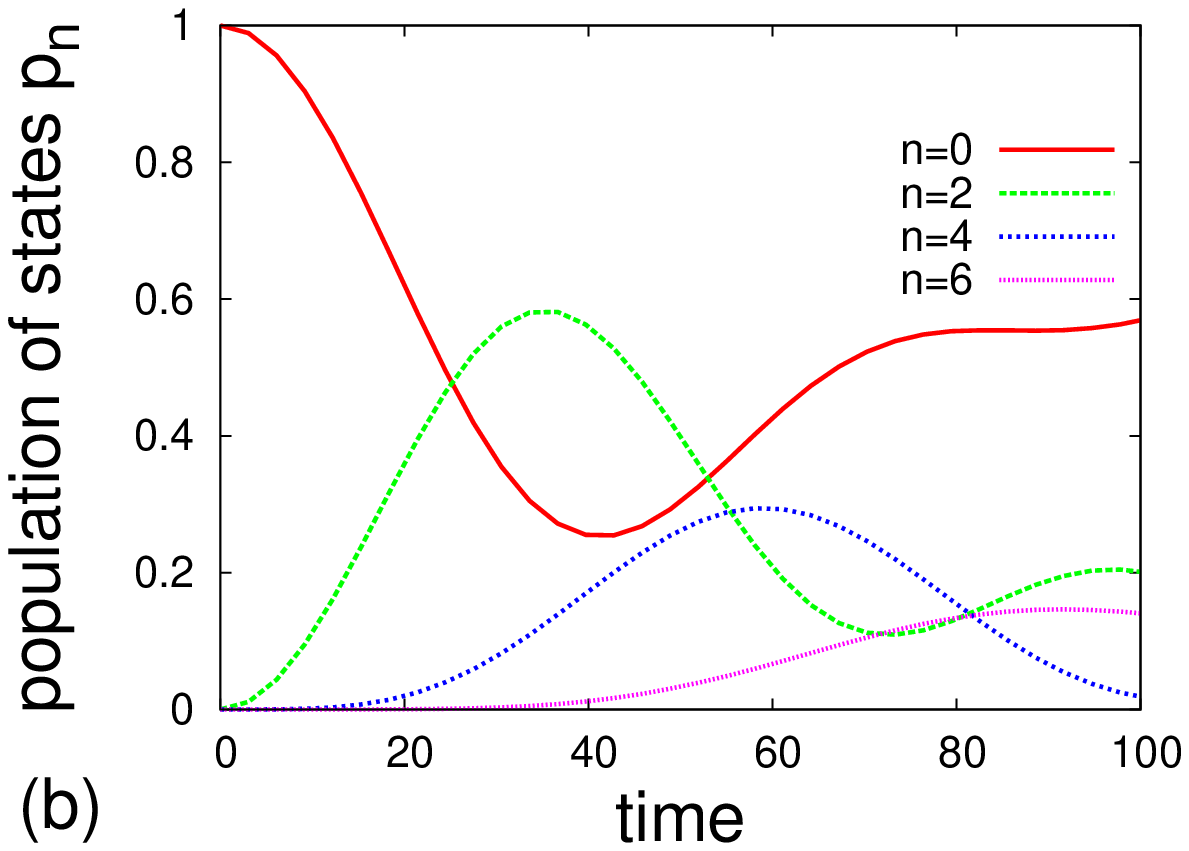}
\caption{Occupation of the adiabatic eigenstates of the Hamiltonian for  $g_0=0$, $\Delta g=0.5$ close to the second and the first resonance (a) $\omega=1.98$ (b) $\omega=0.998$ }
\label{fig3}
\end{figure}

Not only an excitation to the  first excited state of the relative motion is possible, but also to other excited states if the resonant frequency is chosen correspondingly as we can see in Fig.~\ref{fig3} (a). As expected the resonance width though decreases for transitions to  higher states. The effect of the initial $g_0$ and of the amplitude $\Delta g$ is similar to the previous case. 

For larger amplitudes the controllability of the excitation process reduces, as many states are subsequently excited, and simultaneously taking part in a complex time evolution. A typical example is presented in Fig.~\ref{fig3} (b). Still the frequency plays the dominant role and only close to resonances the evolution leads to highly excited states of the spectrum. Mechanisms of acceleration appear then which we will discuss in the following section. We would like to note here that a large $\Delta g$, offer the possibility of 'multi-photon excitations'. In this case even for low frequencies which are an integer ratio of the principle resonant frequency, excitations become possible.  

\section{acceleration via multiple excitation and Floquet analysis}     

We will now more thoroughly examine the case of strong driving, which makes it possible, as we will show, to accelerate the particles, i.e., to increase the mean value of the energy with time. The process of multiple excitations as we have seen, is possible close to resonances, since the spectrum is approximately equidistant. Especially a larger value of $\Delta g$ leads to a covering of wider areas of the energy gaps, and therefore the  comparatively small differences between the gaps effectively drain away. Through this multiple excitation process, the system never returns completely to the ground state, and indeed occupies gradually increasingly higher lying states. This excitation process induces an increase of the energy to very high values  as long as the gaps to higher excited states are in the resonance window. 

\begin{figure}
\includegraphics[width=6.2 cm,height=6.2 cm]{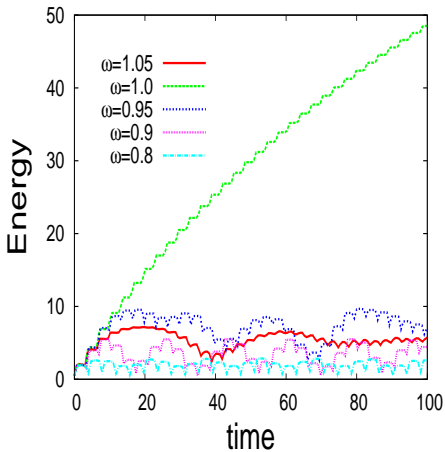}
\caption{Time evolution of the expectation value of the energy for $\Delta g=20.0$, $g_0=0$ and several frequencies close to resonance. Acceleration shown close at the resonant frequency}
\label{fig4}
\end{figure}

We present in Fig.~\ref{fig4}  the time evolution of the expectation value of the energy, close to the first resonance for $\Delta g=20.0$. For values of the frequency sufficiently far from resonance, a repopulation of the ground state in the course of the time can be observed, while a multi-mode behavior is encountered due to an excitation of several states. Approaching the resonance, the instantaneous ground state is never repopulated significantly, on the opposite, higher states of the spectrum are subsequently populated in the same manner as shown in Fig. ~\ref{fig3} (b). This leads to an acceleration, i.e., energy gain of the particles. Our finite time simulations indicate that this energy gain approaches a saturation to very high values of the energy for strong but finite $\Delta g$. 

Let us analyze this resonant mechanism from the perspective of Floquet theory which has been developed for time-periodic Hamiltonians. The Ansatz of the Floquet theory for the time-dependent wave packet reads:
\begin{equation}
\psi_k(r,t)=e^{i q_k t}\sum_n u^k_n(r) e^{in\omega t} 
\end{equation}
where $u^k(r,t)=\sum_n u^k_n(r) e^{in\omega t}$ are the so-called Floquet eigenstates or quasi-energy states, which are time periodic functions expanded here into a Fourier series and $q_k$ are the quasienergies. Introducing this Ansatz into the Schr\"odinger equation for the Hamiltonian in Eq.~(\ref{hamrel}) we find:
\begin{eqnarray}
\label{floqueteq}
 q_k u^k_n (r) &=\left[n\omega+(-\frac{d^2}{dr^2} +\frac{1}{4}r^2) +(g_0+\frac{1}{2}\Delta g)\delta(r) \right]u^k_n (r)
\nonumber\\
          &- \frac{1}{4} \delta(r) \Delta g (u^k_{n-2}(r)+u^k_{n+2}(r))
\end{eqnarray}
The r.h.s. of this equation is the Floquet Hamiltonian which has diagonal, i.e., proportional to $u^k_n$, and off-diagonal terms coupling the different modes $u^k_{n-2}$, $u^k_{n+2}$. To find $u^k_n$ and eigenvalues $q_k$ one should solve this eigenvalue problem by diagonalizing the corresponding Hamiltonian. The harmonic oscillator $f_m(r)$ basis is very convenient to express the Floquet Hamiltonian in our particular problem, not only because of the harmonic part of the potential which has as matrix elements the harmonic oscillator eigenvalues but also because the matrix elements of $\delta(r)$ are of the form $\int{dr f_{m} (r) \delta (r) f_{m'}(r)}= f_m(0)f_{m'}(0)$. An alternative method equally well-suited for our problem is to write the Floquet eigenvectors in terms of parabolic cylinder functions \cite{abramowitz} $D(-\alpha-\frac{1}{2},x)$ which are the solutions of the underlying stationary problem \cite{busch}. In this representation we have 
\begin{equation}
u^k_n(r)=c^k_n D^k_n(-(q_k+n \omega),r)
\end{equation}
with the important properties $D(\alpha,0)=\frac{\sqrt{\pi}}{2^{\frac{1}{2}\alpha+\frac{1}{4}} \Gamma(\frac{3}{4}+\frac{1}{2}\alpha)}$,  $\frac{dD(\alpha,x)}{dx}|_{x=0}=\frac{-\sqrt{\pi}}{2^{\frac{1}{2}\alpha-\frac{1}{4}} \Gamma(\frac{1}{4}+\frac{1}{2}\alpha)}$. By integrating Eq. (\ref{floqueteq}) using these relations we obtain a recursive system of algebraic equations for the coefficients $c_n$. Demanding that this system possesses solutions we obtain the values of quasienergies $q_k$. We have used both methods to derive the eigenspectrum of the Floquet Hamiltonian for our problem and confirmed the agreement crosschecking the results and thereby the numerical convergence.

\begin{figure}
\includegraphics[width=6.2 cm,height=6.2 cm]{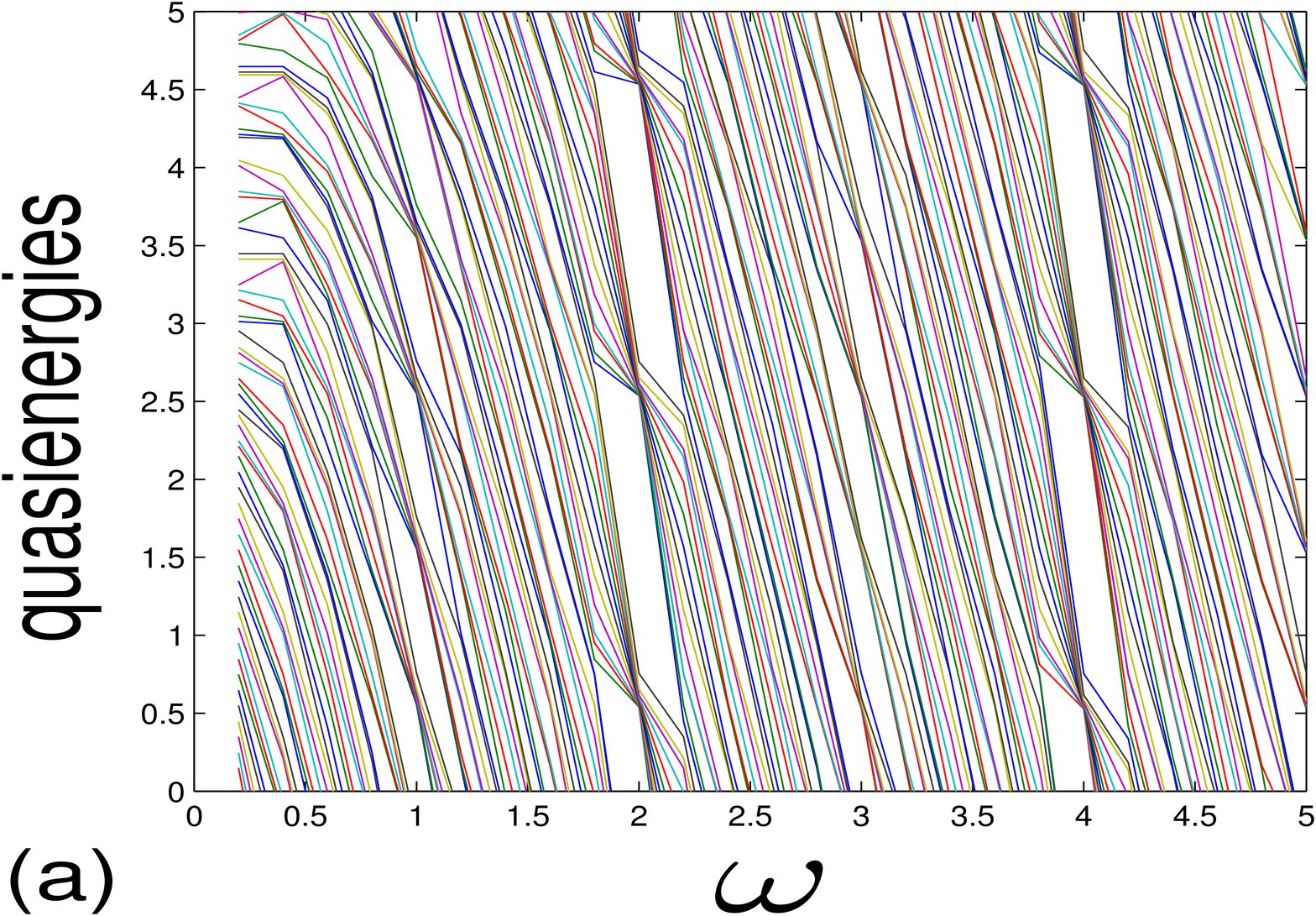}
\includegraphics[width=6.2 cm,height=6.2 cm]{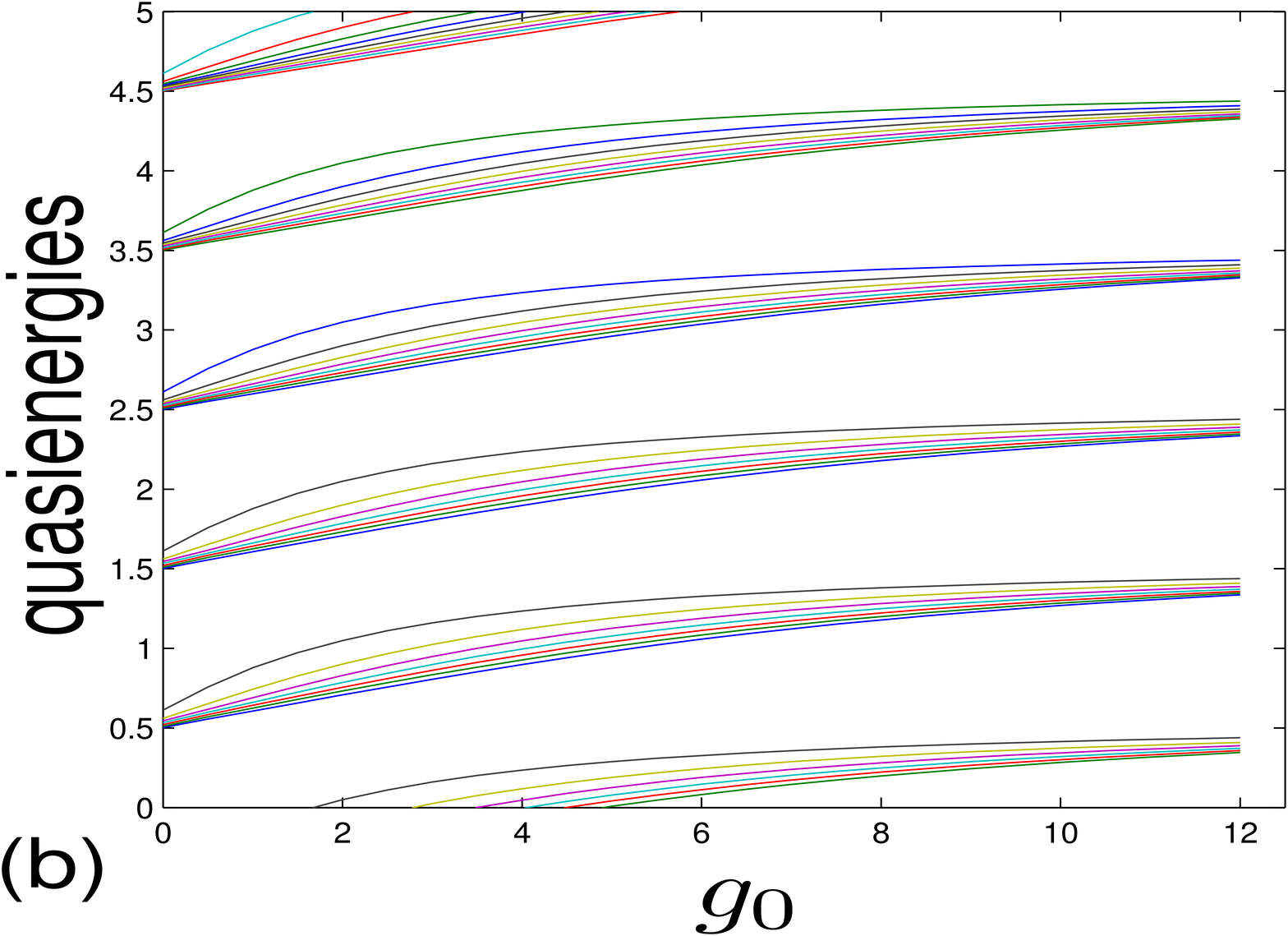}
\caption{Floquet eigenspectrum for (a) increasing driving frequency $\omega$ with $g_0=0.5$ $\Delta g=0.5$, (b) increasing initial interaction strength $g_0$ for the resonant frequency $\omega=1$ and $\Delta g=0.5$}
\label{fig5}
\end{figure}

We show in Fig.~\ref{fig5} (a) the eigenspectrum of the Floquet Hamiltonian, i.e., the Floquet quasi-energies with increasing driving frequency $\omega$  for a typical case $g_0=0.0$  and $\Delta g=1.0$. In the dense spectrum of quasi-energies we encounter points of accumulation when the frequency is at resonant values ($\omega=1,2,3...$). At these points, many Floquet quasi-energies take a value close to the harmonic oscillator eigenenergies ($0.5,1.5,2.5...$), and form Floquet bands. The even-frequency resonances ($\omega=2,4,6...$) correspond to accumulation points with an energy gap $2$, i.e. at the quasi-energy values $0.5,2.5,4.5,...$.  

Let us now show how these results obtained for the Floquet Hamiltonian are connected with the quantum acceleration processes for quantum resonances. In dynamical systems a pure point Floquet spectrum is associated with localized behavior and the energy remains bounded at all times, while singularly continuous components are responsible for diffusive behavior and growth of the energy (see Ref. [\onlinecite{gardiner}] and references therein). At the resonant driving frequencies as we have seen above the quasi-energies are accumulating and approach particular values forming close to continuous areas. This property of the spectrum leads to the acceleration and energy gain.

Besides, in Fig. \ref{fig5} (b) we see that with increasing $g_0$ and for the resonant frequency $\omega=1$ the eigenenergies of the Floquet spectrum deviate from each other, and only come close again as we approach the fermionization limit in the next upper level. This deviation from the accumulation points makes the evolution less diffusive for intermediate interactions, and corresponds to the picture of the instantaneous spectrum, where the energy gaps for intermediate values of $g$ become less equidistant, prohibiting multiple excitation.   

\section{higher atom numbers and finite size effects on collective excitations (breathing mode)}

\begin{figure}
\includegraphics[width=4.2 cm,height=4.2 cm]{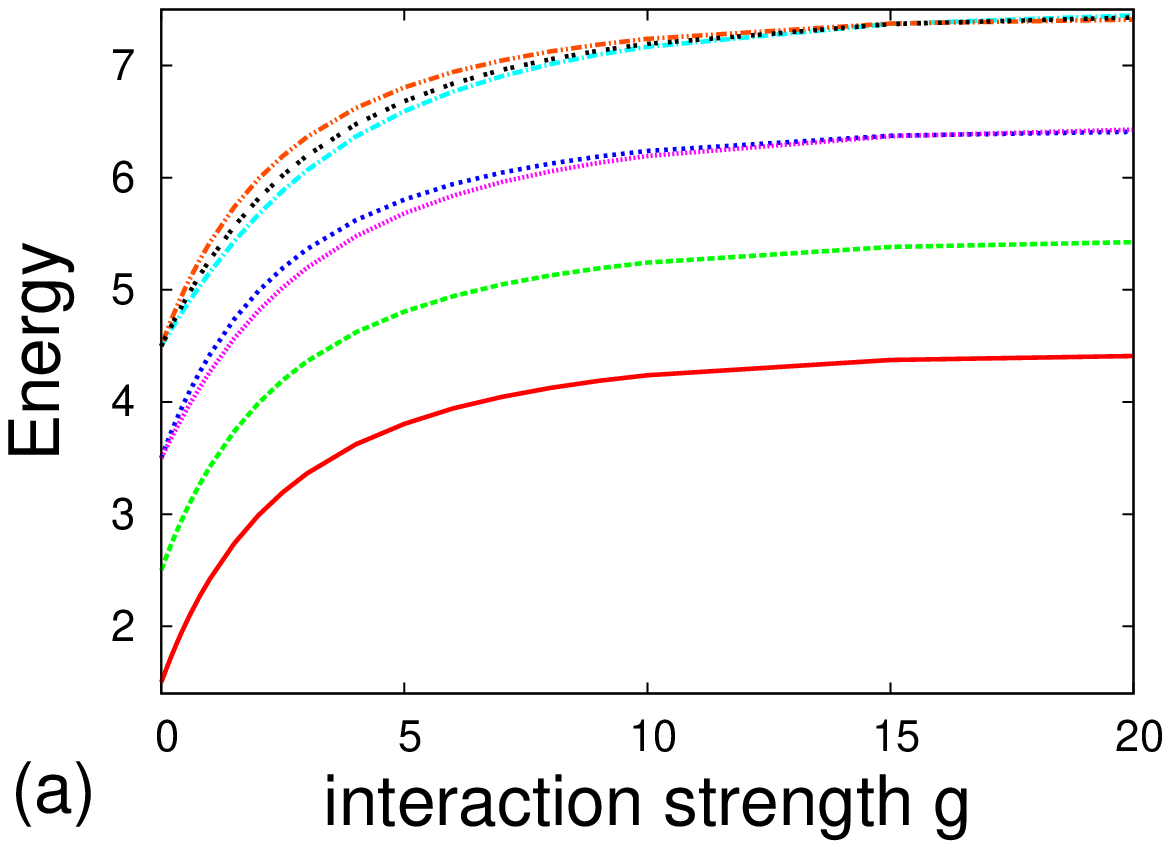}
\includegraphics[width=4.2 cm,height=4.2 cm]{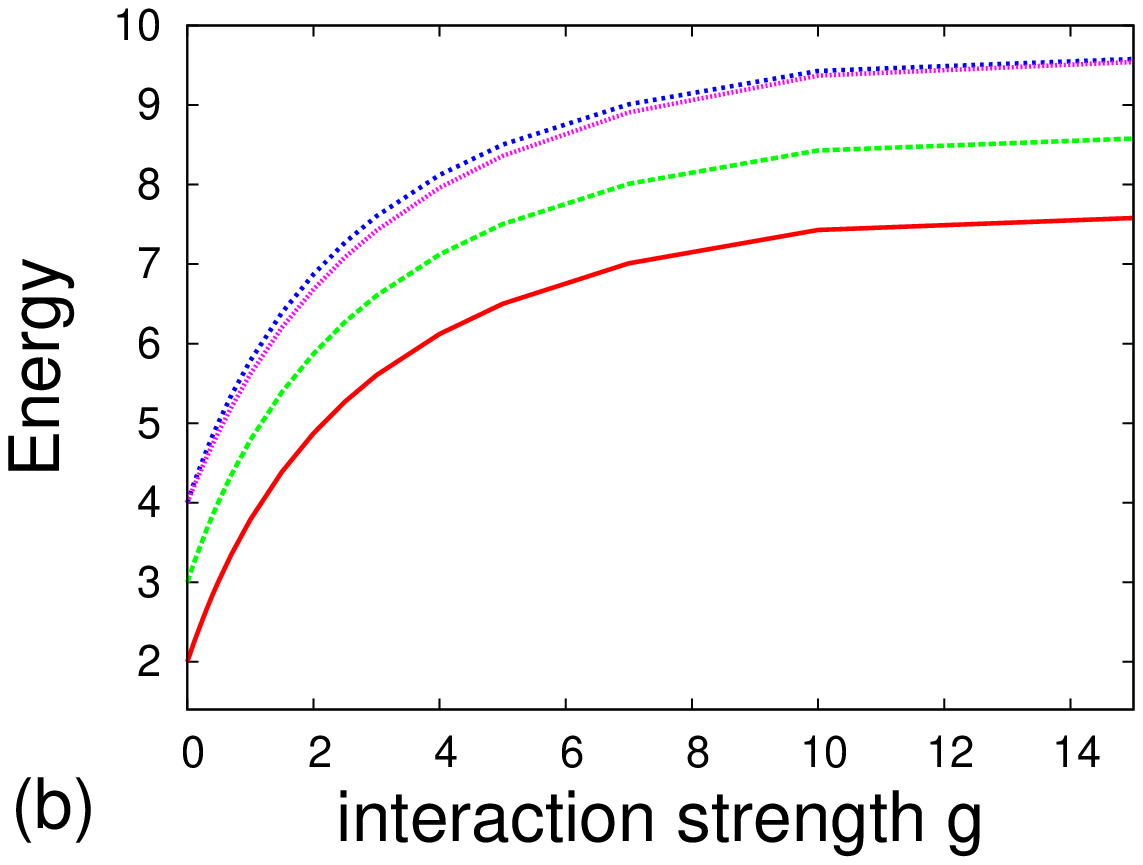}
\caption{The lowest eigenenergies with increasing interaction $g$ for (a) three and (b) four bosons. }
\label{fig6}
\end{figure}

The extensive analysis of the two-body case above, is very useful for an understanding of the effects occurring for higher atom numbers. The main reason for this are the properties of the many body spectrum of the harmonic oscillator including the delta-type interaction. We present in Fig.~\ref{fig6} the energetically low lying part of the spectrum for three and four particles with increasing interaction strength [see also Ref. \onlinecite{sascha}]. We observe that all states show a quite similar evolution and thus the energy gaps between them do not deviate significantly (crossings or anti-crossings do not occur). Also important is that most of the states correspond to excitations of the center of mass, which are not relevant to our study. For example the first excited state and one of the two states of the second excited band which behave exactly like the ground state, correspond to the ground state of the relative motion and to the first and second excitation of the center of mass motion, respectively. 

\begin{figure}
\includegraphics[width=4.2 cm,height=4.2 cm]{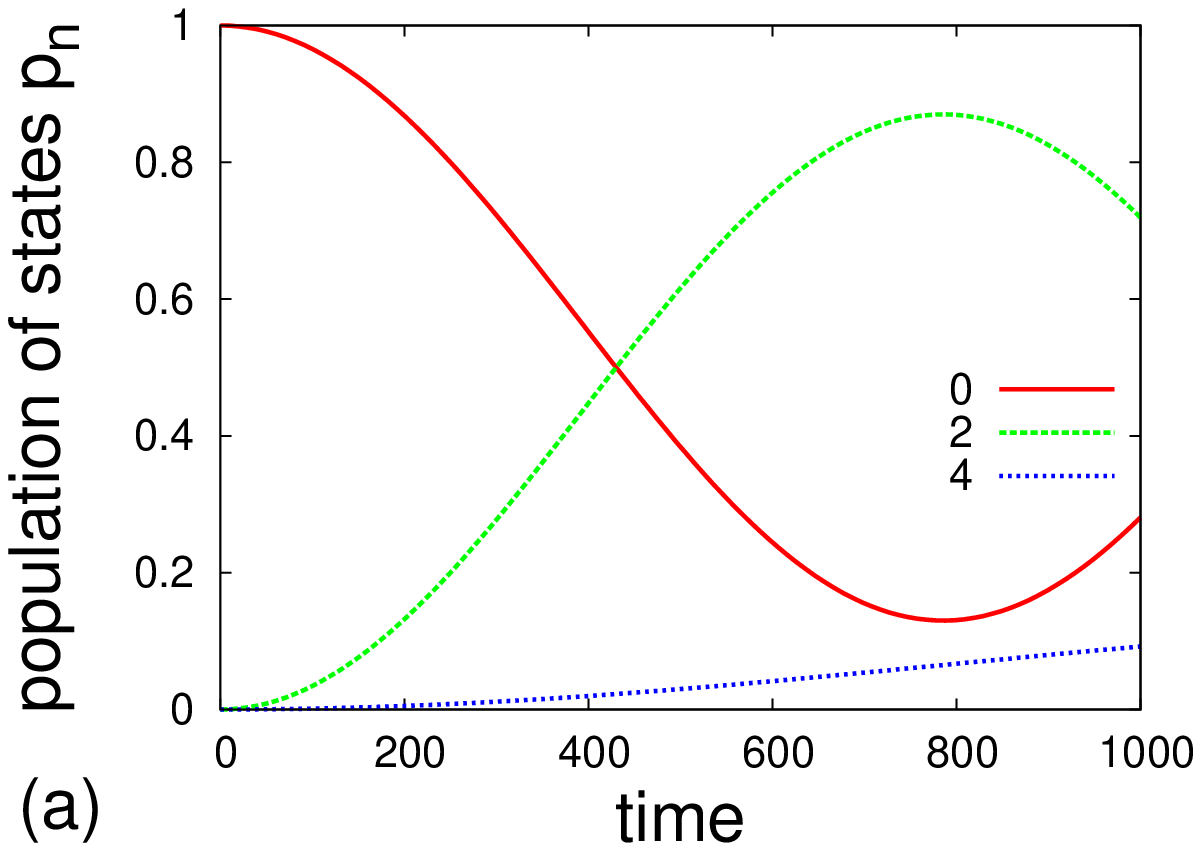}
\includegraphics[width=4.2 cm,height=4.2 cm]{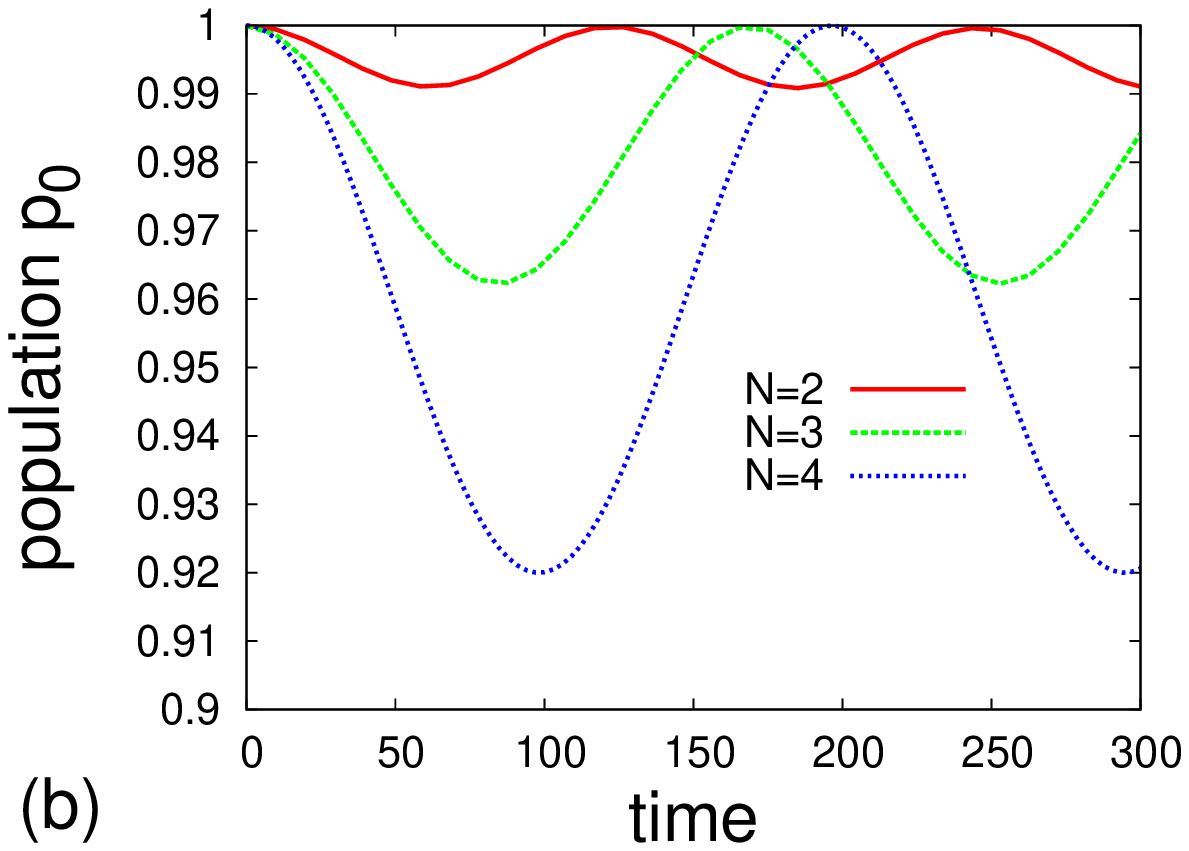}
\caption{ (a) population of instantaneous eigenstates for three particles with $g_0=2.0$, $\Delta g = 0.05$ and $\omega=0.9$. (b) Population of the instantaneous ground state for different numbers of particles $N=2,3,4$ $g_0=0.0$ $\Delta g=1.0$ and $\omega=0.6$}
\label{fig7}
\end{figure}

The time-dependent variation of the interaction strength which affects only the relative motion offers the possibility of a controllable excitation to specific states also for higher atom numbers, since states like those mentioned above which correspond to excitations of the center of mass  do not contribute to the time evolution and are therefore avoided. We demonstrate this in Fig. \ref{fig7} (a) for the case of three particles and an initial value $g_0=2.0$ in the intermediate regime, where an excitation predominantly to the lower state (corresponding to an excitation of the relative motion) of the second excited band of eigenstates is performed [see Fig. \ref{fig6}(a)]. We denote here by the numbers $0,2,4$ the energetically ordered states which correspond to a respective excitation of the many body relative motion. The resonant frequency for excitation from state 0 to state 2 will be referred as the principle resonance in the following.

The above signifies that higher number of particles allow for a similar control of their dynamics as in the case for two particles which is mainly due to the similarities of their underlying energy spectrum. One important difference here is that for equal parameters, systems with larger atom numbers experience a stronger impact of the driving than those with a lower atom number. This can be seen from the the maximum loss of population of the instantaneous ground state with time in Fig. \ref{fig7} (b) in a non-resonant case which gets larger with an increasing number of particles. An explanation for this atom number related effect is the response of the ground state energy to the variation of $g$: 
\begin{equation}
\frac{dE}{dg}\Big\vert_{g=0} = \frac{N(N-1)}{2} \int |\phi_0(x)| ^4dx,
\end{equation}
where we see that the slope of the total energy at $g=0$ increases quadratically with the particle number. Therefore a variation of $g$ possesses a  greater impact for higher atom numbers.  
  
Another important observation concerning the size of the system is the position of the principle resonance in the regime of intermediate interactions. We have seen already that for e.g. $g_0=2.0$ the resonant frequency is lower than for very weakly or strongly interacting initial states. This frequency becomes even lower as the number of particles increase which is based on the decreasing energetical spacing in the corresponding many-body spectra with increasing particle number. Consequently a lower value of frequency is needed for larger atom number for the corresponding resonant excitation via driving of the interaction strength. 

The above observation is important for modes of collective oscillation of the wave function, in analogy to the macroscopic collective oscillations \cite{pethick_smith}. Many measurements in experiments are based on exciting collective modes, which are usually analyzed within the effective mean field descriptions \cite{stringari}. The frequencies of these oscillations can be obtained with a high accuracy from the experimental data usually observing the size or the mean position of the condensate \cite{moritz, haller}, and represent a very important measure for identifying different interaction regimes. For example in one dimension the Thomas Fermi limit, or the Tonks Girardeau and Super-Tonks girardeau gas, possess a very characteristic ratio between the so-called dipole mode which is an oscillation of the center of mass of the condensate with the trap frequency $\omega_d$, and the breathing or first compression mode, where essentially the size of the condensate is oscillating with a frequency $\omega_b$. For the effectively non-interacting limits ($g=0,\infty$) this ratio is $\omega_b/\omega_d=2$ while for the Thomas Fermi limit it is  $\omega_b/\omega_d=\sqrt{3} \approx 1.73 $ \cite{stringari}. These important theoretical results have been confirmed in the corresponding experiments \cite{moritz,haller}.

\begin{figure}
\includegraphics[width=6.2 cm,height=6.2 cm]{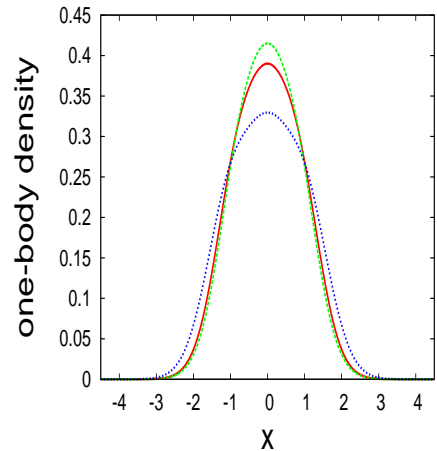}
\caption{One body density for the case of three particles as in Fig. \ref{fig7} for several snapshots.}
\label{fig8}
\end{figure}

From the few body perspective which we examine here, the mode of oscillation that we excite by varying the scattering length, is of a compressional-breathing type. We demonstrate this in Fig. \ref{fig8} where we show several snapshots of the one-body density for characteristic time moments. The excitation to the second state of the spectrum which corresponds to an excitation of the relative motion (and therefore to a broader wave function), possesses therefore the characteristics of a  breathing mode. The characteristic frequency of such a mode, in the intermediate interaction regime could be compared with the mean field result for the Thomas Fermi regime. For non-interacting cases we confirm the ratio $\omega_b/\omega_t=2$. The Thomas Fermi regime applies to large ensembles of particles, and the analogy with a the finite system examined here can be at most indicative. Nevertheless in one-dimensional systems we can take the minimum value of the energy gap which appears close to $g=2.0$ as corresponding to the Thomas-Fermi regime (see also  corresponding arguments in \cite{yiannis}). In doing so, we can reveal finite size corrections for the breathing mode of the Thomas Fermi limit, starting with two particles $\omega_b/\omega_t \approx 1.85$ \cite{privdis} and going to $\omega_b/\omega_t \approx 1.78$ for five particles with a further tendency to decrease with increasing particle number. In the macroscopic limit $N \to \infty$ we should approach the mean field estimate $\omega_b/\omega_t=\sqrt{3} \approx 1.73 $ \cite{stringari}.

This interrelation of the collective mode frequencies and their finite size deviations within the few-body spectrum are probably valuable for further studies on connecting, checking and reinterpreting mean field results in the light of the exact many-body spectrum and behaviour.

\section{Conclusion}

We examined the effects of a periodically driven interaction strength on ultracold bosonic systems in a one-dimensional harmonic trap which can be realized by time a modulation of magnetic fields utilizing Feshbach resonances or periodically changing the transversal confinement length in a waveguide via laser fields. We have shown that the feature of near equidistant energy levels for arbitrary interaction strength  for the relative motion  of two particles in a harmonic trap but also for the corresponding many-body spectrum for larger atom numbers  has important consequences for the excitation dynamics. In particular, the energetical spacing yields resonant driving frequencies, which one can employ to excite particular states of bosonic relative motion. We underline that unlike other driven many-body systems, the variation of the interaction strength offers the possibility to design excitations of the relative motion of a certain species exclusively with a very high degree of controllability for certain regimes of the driving parameters. Approaching the resonant frequencies the atoms excite to the corresponding excited level of the relative motion which has been demonstrated here by calculating the occupation of instantaneous eigenstates. For strong driving amplitudes in the vicinity of the resonances the energy reaches out to very high values with several and successive excitations to energetically higher lying states of the spectrum. This multi-excitation process of acceleration is also analyzed via the properties of the Floquet spectrum. The initial interaction strength distorts the energy spectrum shifting the position of the resonances, while highly correlated initial states are quite insensitive with respect to the changes of the repulsive interaction. We have shown via our exact numerical calculations, that for any number of particles there is a similar and for larger ensembles even more sensitive response to the driving of the interaction strength leading to higher excitation amplitudes. Effects due to the finite size of the system, are also analyzed from the perspective of collective oscillation modes, and especially the analog to the macroscopic breathing mode is established, thereby discussing similarities and deviations from mean field approaches. Especially the two-body problem discussed here represents a case of interest to the experiment of distinguishable fermions in a harmonic trap \cite{friedhelm,selim}.  Interesting outlooks are the exploration of different  driving laws, including the possibility to alter between repulsive and attractive interactions or using different potential landscapes.

\acknowledgments

The authors acknowledge many fruitful discussions with F.K. Diakonos concerning the Floquet analysis and thank Hans-Dieter Meyer for helpful discussions and comments. Financial support by the Deutsche Forschungsgemeinschaft is gratefully acknowledged. 

\appendix
\section{Computational Method}

Treating time-dependent Hamiltonians with many interacting degrees of freedom is a computationally very demanding problem. In this work for all numerical calculations we rely on the Multi-Configurational Time-Dependent Hartree (MCTDH) method \cite{mctdhbook,meyer90,beck00},  primarily a wave-packet dynamical tool known for its outstanding efficiency in high-dimensional applications. 
The underlying idea of MCTDH is to solve the time-dependent Schr\"odinger equation

\begin{equation}
\left\{ \begin{array}{c}
i\dot{\Psi}=H\Psi\\
\Psi(Q,0)=\Psi_{0}(Q)\end{array}\right.\label{eq:TDSE}
\end{equation}
as an initial-value problem by an expansion in terms of direct (or Hartree)products $\Phi_{J}$:

\begin{eqnarray}
\nonumber
\Psi(Q,t)&=&\sum_{J}A_{J}(t)\Phi_{J}(Q,t)
\nonumber\\
&\equiv&\sum_{j_{1}=1}^{n_{1}}\ldots\sum_{j_{f}=1}^{n_{f}}A_{j_{1}\ldots j_{f}}(t)\prod_{\kappa=1}^{f}\varphi_{j_{\kappa}}^{(\kappa)}(x_{\kappa},t),\label{eq:mctdh-ansatz}
\end{eqnarray}
using a convenient multi-index notation for the configurations, $J=(j_{1}\dots j_{f})$, where $f=N$ denotes the number of degrees of freedom and $Q\equiv(x_{1},\dots,x_{f})^{T}$. The \emph{single-particle functions} $\varphi_{j_{\kappa}}^{(\kappa)}$ are in turn represented in a fixed, primitive basis implemented on a grid. For indistinguishable particles as in our case, the single-particle functions for each degree of freedom $\kappa=1,\dots,N$ are of course identical in both type and number ($\varphi_{j_{\kappa}}$, with $j_{\kappa}\le n$).

In the above expansion, both the coefficients $A_{J}$ and the Hartree products $\Phi_{J}$ are time-dependent. Using the Dirac-Frenkel variational principle, one can derive equations of motion for both $A_{J},\Phi_{J}$. This conceptual complication offers an enormous advantage: the basis $\{\Phi_{J}(\cdot,t)\}$ is variationally optimal at each time $t$, allowing us to keep it fairly small. Exactly for this reason MCTDH is ideally designed to solve time-dependent Hamiltonians like driven systems as the one we tackle here. The permutation symmetry can be enforced by symmetrizing the coefficients $A_{J}$, but the ground state is automatically of bosonic character. 

In addition the Heidelberg MCTDH package  \cite{mctdh:package} incorporates the so-called \emph{relaxation method} which provides a way to obtain the lowest \emph{eigenstates} of the system by propagating some wave function $\Psi_{0}$ by the non-unitary $e^{-H\tau}$ (\emph{propagation in imaginary time}.) As $\tau\to\infty$, this automatically damps out any contribution but that stemming from the true ground state, \[ e^{-H\tau}\Psi_{0}=\sum_{J}e^{-E_{J}\tau}|J\rangle\langle J|\Psi_{0}\rangle.\] In practice, one relies on a more sophisticated scheme termed \emph{improved relaxation} \cite{meyer03}. Here $\langle\Psi|H-E|\Psi\rangle$ is minimised with respect to both the coefficients $A_{J}$ and the configurations $\Phi_{J}$. The equations of motion are solved iteratively, first for $A_{J}(t)$ (by diagonalisation of $\langle\Phi_{J}|H|\Phi_{K}\rangle$ with fixed $\Phi_{J}$) and then propagating $\Phi_{J}$ in imaginary time over a short period. The cycle will then be repeated. This method is used in the present work to compute the eigenstates and spectrum for $N=3,4,5$ particles.

As it stands, the effort of this method scales exponentially with the number of degrees of freedom, $n^{N}$. Just as an illustration, using $15$ orbitals and $N=5$ requires $7.6\cdot10^{5}$ configurations $J$. This restricts our analysis in the current setup to about $N=O(10)$, depending on how decisive correlation effects are. Therefore we are limited in terms of numerical convergence to the number of particles we can treat for large variations of the interaction strength. The number of orbitals needed for a time-evolution of our driven system also depends on how prominent the excitations are, i.e., how close to resonance the driving frequency is. The latter fact sets limitations to examining the long-time dynamics where the behaviour involves many excitations rather than oscillatory. By contrast, the dependence on the primitive basis, and thus on the grid points, is not as severe. In our case, the grid spacing should of course be small enough to sample the interaction potential. The length of the grid should be also choosen sufficiently large especially in cases where the driving frequency  approaches resonances and the time-dependent wave function covers a very wide coordinate range.


\begin{thebibliography}{10}
\bibitem{privdis}
Private communication: The experiment in Heidelberg \cite{friedhelm,selim} has measured up to $\omega_b/\omega_t=1.89$ for the excitation of the two-body relative motion.
\end{thebibliography}
\end{document}